\documentstyle[12pt,epsf]{article}
\begin{document}
\thispagestyle{empty}
\rightline{KIAS-P00046}
\rightline{UOSTP-00105}
\rightline{{\tt hep-th/0007107}}

\

\def\tr{{\rm tr}\,} \newcommand{\beq}{\begin{equation}}
\newcommand{\eeq}{\end{equation}} \newcommand{\beqn}{\begin{eqnarray}}
\newcommand{\eeqn}{\end{eqnarray}} \newcommand{\bde}{{\bf e}}
\newcommand{\balpha}{{\mbox{\boldmath $\alpha$}}}
\newcommand{\bsalpha}{{\mbox{\boldmath $\scriptstyle\alpha$}}}
\newcommand{\betabf}{{\mbox{\boldmath $\beta$}}}
\newcommand{\bgamma}{{\mbox{\boldmath $\gamma$}}}
\newcommand{\bbeta}{{\mbox{\boldmath $\scriptstyle\beta$}}}
\newcommand{\lambdabf}{{\mbox{\boldmath $\lambda$}}}
\newcommand{\bphi}{{\mbox{\boldmath $\phi$}}}
\newcommand{\bslambda}{{\mbox{\boldmath $\scriptstyle\lambda$}}}
\newcommand{\ggg}{{\boldmath \gamma}} \newcommand{\ddd}{{\boldmath
\delta}} \newcommand{\mmm}{{\boldmath \mu}}
\newcommand{\nnn}{{\boldmath \nu}}
\newcommand{\diag}{{\rm diag}}
\newcommand{\bra}[1]{\langle {#1}|}
\newcommand{\ket}[1]{|{#1}\rangle}
\newcommand{\sn}{{\rm sn}}
\newcommand{\cn}{{\rm cn}}
\newcommand{\dn}{{\rm dn}}
\newcommand{\tA}{{\tilde{A}}}
\newcommand{\tphi}{{\tilde\phi}}
\newcommand{\bpartial}{{\bar\partial}}
\newcommand{\br}{{{\bf r}}}
\newcommand{\bx}{{{\bf x}}}
\newcommand{\bk}{{{\bf k}}}
\newcommand{\bq}{{{\bf q}}}
\newcommand{\bQ}{{{\bf Q}}}
\newcommand{\bp}{{{\bf p}}}
\newcommand{\bP}{{{\bf P}}}
\newcommand{\thet}{{{\theta}}}
\renewcommand{\thefootnote}{\fnsymbol{footnote}}
\

\vskip 0cm
\centerline{ \Large\bf Elongation of  Moving Noncommutative Solitons }

\vskip .2cm

\vskip 1.2cm
\centerline{ 
Dongsu Bak,$^a$\footnote{Electronic Mail: dsbak@mach.uos.ac.kr} 
and Kimyeong Lee $^{b}$\footnote{Electronic Mail: 
klee@kias.re.kr} 
}
\vskip 10mm 
\centerline{ \it $^a$ Physics Department, 
University of Seoul, Seoul 130-743, Korea} 
\vskip 3mm 
\centerline{ \it $^b$ School of Physics, Korea Institute for Advanced 
Study} 
\centerline{ \it 207-43, Cheongryangryi-Dong, Dongdaemun-Gu, Seoul 
130-012, Korea 
} 
\vskip 1.2cm 

%\vskip 1.2cm

\begin{quote}
{%\baselineskip 16pt 
We discuss the characteristic properties of
noncommutative solitons moving with constant velocity.  As
noncommutativity breaks the Lorentz symmetry, the shape of moving
solitons is affected not just by the Lorentz contraction along the
velocity direction, but also sometimes by additional `elongation'
transverse to the velocity direction. We explore this in two examples:
noncommutative solitons in a scalar field theory on two spatial
dimension and `long stick' shaped noncommutative U(2) magnetic
monopoles. However the elongation factors of these two cases are
different, and so not universal.} 
\end{quote}
%\centerline{\today}

%\pacs{14.80.Hv,11.27.+d,14.40.-n}

\newpage

Solitons in the noncommutative field theories have attracted much
attention recently\cite{nekrasov, hashimoto, bak,
hata,strominger,harvey, moriyama, goto, gross}.  Localized solitons in
a noncommutative scalar theory of spatial dimensions higher than one
is already peculiar because they lost their identity in the
commutative case as dictated by the Hobart-Derrick
theorem\cite{strominger,hobart,derrick}.  On the other hand, monopoles or
dyons of the commutative super Yang-Mills (SYM) theories become a
sticklike.  From the D-brane picture, D-strings connecting D3 branes
become tilted in noncommutative case. In the field theory picture, the
image of a D-string on three space appears as a finite segment of
Dirac string, whose two ends are like Dirac magnetic monopoles of two
different $U(1)$ subgroups of
$U(2)$\cite{hashimoto,bak,hata,moriyama,goto,gross}.

The detailed dynamical aspects of these noncommutative solitons are
not much pursued.  Investigation of the free motion of one
noncommutative soliton will be the first step toward the understanding
of their solitonic moduli dynamics. (See Ref.~\cite{yi} for the the
moduli space of a single caloron made of $N$ monopoles in $U(N)$
theory on noncommutative $R^3\times S^1$.)

In this note, we consider free motions of a noncommutative soliton, which
are not trivial because the systems lack the Lorentz invariance. The
change of the shape of the soliton, for example, is not just dictated
by the Lorentz contraction but further deformation is induced since
the effective noncommutativity scale is changed due to the structure
of $*$-product. The key finding is that these solitons can be elongated
along transverse to the velocity direction, but the elongation effect
is not universal.

For the solitons of two dimensional scalar theory, the size transverse
to the motion is elongated by the factor $\sqrt{\gamma}$ while the
longitudinal size becomes contracted by the factor $1/\sqrt{\gamma}$,
preserving the area size of the soliton, where the Lorentz contraction
factor is $\gamma^{-1} = \sqrt{1-v^2}$.  In particular, when the
velocity approaches the light velocity, the noncommutative soliton
looks like a very long and thin string stretched in the transverse
direction of the motion.

In the case of $U(2)$ BPS monopoles interpreted as tilted D-strings
connecting two parallel $D3$ branes, the tilting is affected by the
motion. Equally, the length of the Dirac string connecting two different
$U(1)$ monopoles is affected. When the direction of motion is transverse to the
Dirac string, the string length gets elongated by the factor $\gamma$. When the
direction of motion is parallel to the string, the length is contracted by
the factor $\gamma^{-1}$. As a by product, we get 
tilting and the tension of static $(p_e,q_m)$- dyons in similar
perspective.  

\noindent{\large \sl Noncommutative solitons in (2+1) dimensional
scalar field theory}

We shall first consider the noncommutative soliton arising
in 2+1 dimensional scalar theory,
\begin{equation}
L=\int d^2x\left( {1\over 2}\partial_\mu\phi \partial^\mu\phi 
 -V(\phi)
\right)\,,
\label{lag}
\end{equation} 
where the product between  fields is defined by 
the $*$-product, 
%(Moyal product),
\begin{equation}
a(\bx)* b(\bx)\equiv \Bigl(e^{{i\over 2}\theta\epsilon^{ij} 
\partial_i \partial'_j} a(\bx) b(\bx')\Bigr){\Big\vert}_{\bx=\bx'}\,.
\label{star}
\end{equation}
%with $\partial\wedge 
%\partial'\equiv $.
For simplicity, we shall consider the case where
 the potential has its absolute minimum at $V(0)=0$ and 
the other local minimum at some $\lambda$ with $V(\lambda) > 0$.
As shown in detail in Ref.~\cite{strominger}, the localized 
soliton
corresponds to a false vacuum bubble where the noncommutativity
prevents its collapse to a zero size. This is contrasted to the
case of commutative scalar field theory where localized solitons
do not exist at all as dictated by the Hobart-Derrick 
theorem\cite{hobart,derrick}. 

The static localized  noncommutative soliton solution will satisfy
\begin{equation}
 -\nabla^2\phi+ V'(\phi)=0\,.
\label{solequation}
\end{equation}
In the large $\theta$ limit, the potential part is dominant
over the kinetic contributions.
This can be easily shown by introducing dimensionless coordinates
$\tilde{x}_i=x_i/\sqrt{\theta}$. The energy functional is written 
as
\begin{equation}
E=\int d^2\tilde{x}\,\left({1\over 2} |\tilde\nabla \phi|^2
+\theta V(\phi)\right)\,.
\label{energy}
\end{equation}
where the $*$-product is defined in terms of 
$\tilde{x}$ with $\theta=1$. Thus we clearly see that
the kinetic contribution is negligible compared to the
 potential contribution. Neglecting the kinetic term, 
the static normalizable solution in this limit 
can be constructed with help of 
projection 
function,
\begin{equation}
P_n({\bf r})=2 (-1)^n  e^{-{r^2\over \theta}} L_n(2 r^2/\theta)
\label{projection}
\end{equation}
where $L_n(x)$ is the n-th order Laguerre polynomial.
These functions work as projection operators under 
the 
%Moyal 
$*$-product; they satisfy $P_n * P_m =\delta_{nm} P_n$\cite{fairlie}.
The most 
general radially symmetric  normalizable solutions  
of $V'(\phi)=0$ are then 
given by  
\begin{equation}
\phi=\sum_n w_n P_n\,,
\label{nonsol}
\end{equation}
where $w_n$ belongs to the set $\{\lambda_l\}$ of real extrema
of $V(x)$. Let us, for example, take the simplest 
solution,
\begin{equation}
\phi(x,y)=\lambda P_0(x,y)=
2 \lambda  e^{-r^2/\theta}\,.
\label{nonsol1}
\end{equation}
The size of the soliton is approximately $R=\sqrt\theta$. By
the axial symmetry, x-directional size $L_x=\sqrt\theta$ is
the same as the y-directional size $L_y=\sqrt\theta$. 
Let us call their potential and kinetic energy to be
\begin{equation}
K_0=\int d^2\tilde{x} \,\frac{1}{2} (\tilde{\nabla}\phi)^2,\;\;\; 
U_0= \theta \int d^2\tilde{x} \, V(\phi)
\end{equation}
Their order of magnitude is $K_0\sim {\cal O}(1)$ and $U_0\sim {\cal
O}(\theta).$ Thus the rest mass, $E_0= K_0+ U_0$, is dominated by the
potential.

Now let us consider any static solution $\bar\phi(x,y;\tilde\theta)$
of (\ref{solequation}) with a noncommutativity scale
$\tilde\theta$. Due to the rotational symmetry, we consider just a
soliton moving along the x axis. In the commutative case, the time
dependent solution describing a moving profile can be constructed by
boosting the static soliton by
\begin{eqnarray}
 t'=\gamma(t- vx)\,, \ \  x'=\gamma (x-vt)\,,\ \ y'=y\,, 
\label{boost}
\end{eqnarray}
with $\gamma=1/\sqrt{1-v^2}$.
Because the 
symmetry under the Lorentz boost is explicitly 
broken by the 
noncommutativity, the Lorentz boost no longer generates
a new solution. 
%Instead, noting the fact,
%\begin{eqnarray}
%\gamma\theta{\partial \over \partial x'} f(x',y') 
%{\partial \over \partial y'} g(x',y')
%=\theta{\partial \over \partial x}f(\gamma(x-vt),y) 
%{\partial \over \partial y} 
%g(\gamma(x-vt),y)\,,
%\label{moyal}
%\end{eqnarray}
%we conclude that 
Instead, the moving solution is given by
\begin{equation}
\phi_v(x,y,t;\theta)=\bar\phi(\gamma(x-vt),y;\gamma\theta)\,,
\label{moving}
\end{equation}
which satisfies the equation of motion,
\begin{equation}
\partial_t^2\phi_v-\nabla^2\phi_v+V'(\phi_v)=0\,.
\label{moving1}
\end{equation}
Namely the solution is obtained not by a simple Lorentz boost but
by the boost accompanied by rescaling of  $\theta$.

%The solution is no 
%longer axially symmetric. 
The 
deformation of the shape is not just a conventional Lorentz contraction
because the effective noncommutative scale 
$\theta_{\rm eff}$ is now $\gamma\theta$. 
For a given instant of time $t$,
the size in each direction is now
\begin{eqnarray}
L'_x(v)={\sqrt{\theta}\over \sqrt{\gamma}}={L_x\over \sqrt{\gamma}},
\ \  L'_y (v)=\sqrt{\gamma}\,\sqrt{\theta}={\sqrt{\gamma} L_y}\,.
\label{primesize}
\end{eqnarray} 
It is interesting to note that the area size of the 
noncommutative soliton
is preserved as
\begin{eqnarray}
A(v)=L'_x(v) L'_y (v)= L_x L_y\,.
\label{area}
\end{eqnarray}
This area preserving character is consistent with the fact that
the uncertainty relation set by the noncommutativity of
the coordinate $\Delta x\Delta y \sim \theta$. As argued in 
Ref.~{\cite{strominger}}, this uncertainty relation is responsible 
for the size of the soliton, without which the soliton would 
collapse to zero size. 
As $v$ grows, the transverse size to the motion 
grows as $\sqrt{\gamma}$, reflecting another UV/IR mixing of
the noncommutative field theory. However the growth differs
from those observed  in the wave function of quantum bound 
state\cite{yee}
or in the dipole nature described in Ref.~\cite{susskind}. 
The velocity dependence  of the size is illustrated in Fig. 1.
\begin{figure}[tb]
\epsfxsize=4.7in
%\centerline{
\hspace{.8in}\epsffile{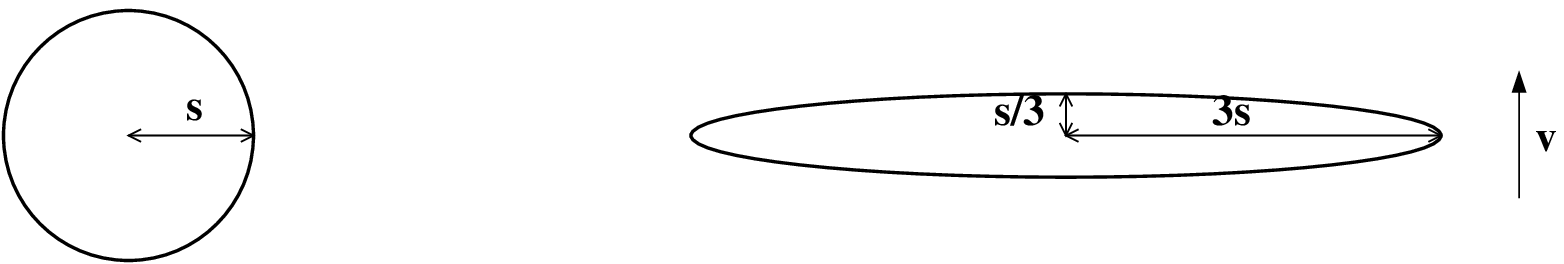}
%}
%\vspace{.1in}
\\
{\small Figure~1:~The static soliton with $s\equiv \sqrt{\theta}$
and the shape of moving soliton with $\gamma=9$.}
\end{figure}

For large $\theta$, the kinetic energy contribution to the energy
 is still of O(1) in $\theta$ as
\begin{eqnarray}
K(v)= {1\over 2} \int d^2 x\, (|\dot\phi|^2+ |\nabla\phi|^2)
= {1\over 2\gamma} \int d^2 x'\,  \left(\gamma^2 (1+ v^2 )
|\partial_{x'}\phi|^2+ |\partial_{y'}\phi|^2\right)= \gamma K_0
\label{kinetic}
\end{eqnarray} 
where we have used the fact that the soliton at the rest frame is
rotationally symmetric.  Thus the potential energy contribution is
dominant over the kinetic part and, consequently, the potential energy 
is given by
\begin{eqnarray}
U(v)= \int d^2 x V(\phi (\gamma(x-vt), y;\gamma\theta)) = {1\over
\gamma}\int d^2 x' V(\phi(x',y';\theta\gamma)) =U_0\,.
\label{kinetic1}
\end{eqnarray} 
Here we have used the fact that $U_0$ is linear in $\theta$. Thus the
total energy transforms as
\begin{equation}
E(v) = \frac{1}{\sqrt{1-v^2}} K_0 + U_0\,.
\end{equation}
In the large $\theta$ limit, the potential energy is dominant until
the velocity is highly relativistic so that $v\sim
\sqrt{1-(K_0/U_0)^2}$.  In case of ordinary solitons in 1+1
dimensional sine-Gordon model or monopoles in SYM theories, the energy
scales as those of ordinary massive particles; $E(v)=\gamma E_0$.
Thus, the behavior of energy of the noncommutative solitons is again
quite different from that of the conventional soliton.  We now turn to
the case of momentum of the moving soliton.  Using the translational
invariance of the system, the conserved momentum may be constructed
using the Noether procedure and the resulting expression reads,
\begin{eqnarray}
{\bf P}=\int d^2 x \, \partial_t \phi \nabla \phi\,.
\label{momentum}
\end{eqnarray} 
The momentum of the noncommutative soliton is then evaluated as
\begin{eqnarray}
P_x=\gamma v\int d^2 x' \left({\partial\phi\over \partial
{x'}}\right)^2 = v\gamma K_0
\label{momentum2}
\end{eqnarray} 
Hence the momentum is not given by $\gamma v E_0$ but its value
is much smaller compared to a particle with rest mass $E_0$.

Because of the change of the shape of moving noncommutative solitons,
the characteristic of classical scattering, for example, ought to
differ from ordinary particles with short-ranged interactions.  As the
relative velocity grows, the size felt becomes bigger, and the cross
section is expected to grow, though a detailed analysis is necessary
to see this effect explicitly.

\noindent{\large\sl Noncommutative U(2) monopole and $(p_e,q_m)$-dyons}

%The U(2) monopole solution is recently discussed in many literatures. 
%Although the explicit solution is not given for the 
%noncommutative U(2) monopole, the commutative description is given 
%explicitly in Ref.~\cite{moriyama}. We shall not need the detailed form 
%of the solution. 

We begin by recapitulating the static properties of noncommutative
monopole in the N=4 supersymmetric Yang-Mills theory.  We shall
restrict our discussion to the case of $U(2)$ gauge group.  Among the
six Higgs fields, only a Higgs field $\phi$ plays a role in the
following discussion.  The bosonic part of the action is then given by
\begin{equation}
S= \frac{1}{g^2_{\rm YM}} \int d^4x\; {\rm tr} \Bigl( -\frac{1}{4}F_{\mu\nu} 
 F^{\mu\nu}+ \frac{1}{2} D_\mu\phi  D^\mu\phi \Bigr).
\label{lag2}
\end{equation}
We shall 
take the only nonvanishing components to be 
$\theta_{12}=-\theta_{21}\equiv\theta$.
%$F_{\mu\nu}$ and $D_\mu\phi$
%are defined respectively by
%\begin{eqnarray}
%&&F_{\mu\nu}\equiv \partial_\mu A_\nu-\partial_\nu A_\mu
%+i(A_\mu * A_\nu-A_\nu * A_\mu)\nonumber\\ 
%&&D_{\mu}\phi\equiv \partial_\mu \phi 
%+i(A_\mu * \phi-\phi *A_\mu)
%\label{fieldstrength}
%\end{eqnarray}
The four vector potential and $\phi$ 
belong to $U(2)$ Lie algebra generated by
$%T_{4}=
{1\over {2}}I_{2\times 2}$ and 
$%(T_1,T_2,T_3)=
{1\over {2}}(\sigma_1,\sigma_2,\sigma_3)$.
% normalized 
%by  $\tr T_m T_n={1\over 2}\delta_{mn}$. 
We set the vacuum 
expectation value of the Higgs field
in the asymptotic region 
as $
%\langle \phi\rangle= 
u\sigma_3/2$.

The energy functional 
\begin{eqnarray}
E\!=\! {1\over 2 g^2_{\rm YM}}\!\!
\int\! d^3x \,\tr \Bigl(E^2 + |D_0\phi|^2  
+ B^2+|D\phi|^2 \Bigr) \ge {1\over g^2_{YM}}
\int_{r=\infty}\!\! dS_k \tr B_k\phi,   
\label{energy2}
\end{eqnarray}
is bounded as in the case of the ordinary supersymmetric Yang-Mills theory.
 The bound is saturated if
the  BPS equation 
\begin{eqnarray}
B=D\phi\,.
\label{bps}
\end{eqnarray}
%is satisfied. 
The mass for the monopole solution  is then
\begin{eqnarray}
M= {2\pi q_m \over g^2_{\rm YM}} u
\label{mass}
\end{eqnarray}
where we define the magnetic charge $Q_M$  by
\begin{eqnarray}
q_m={1\over 2\pi u} \int_{r=\infty} dS_k \; \tr B_k \phi.
\label{charge}
\end{eqnarray} 
The charge is
 to be quantized at integer values even 
in the noncommutative case. This is because the fields in the 
asymptotic region are slowly varying and, hence, the standard
argument of the topological quantization of the magnetic charge holds.
%in the noncommutative theory. 
%Below, we shall restrict our discussion
%for the case of $Q_M=1$.

The $U(2)$ noncommutativity monopole solution has been investigated in
Ref.~\cite{hashimoto,bak,hata, goto} and the solution to the
second order in $\theta$ has been found.  In Ref.~\cite{goto} the
full brane configuration in the commutative SYM picture was found,
which is related to the noncommutative description via the
Seiberg-Witten map\cite{seiberg}.  The monopole (D-string) is tilted
between two parallel D3 branes as schematically illustrated in Fig. 2,
where we interpret $\phi \, l_s^2 $ as the transverse coordinate
$X_4$.  The configuration has an axial symmetry along z-axis and the
projected image to the three space has the z-directional size
$L_z=u\theta$. The gauge symmetry $U(2)$ is spontaneously broken to
$U(1)\times U(1)'$. The two points where D string meet two D3 branes
are like $U(1)$ monopole and $U(1)'$ anti-monopole, which are now
separated in finite interval of length $u\theta$ along the z direction
and connected by the Dirac string of finite tension. This Dirac string
is the  image of the tilted D-string on three space.

%
%
%The scale relevant for the image 
% can be directly seen from the Nahm's
%construction of the noncommutative monopole. The kernel
%equation\cite{bak} is given by
%\begin{eqnarray}
%-{d\over d(s U)}V + \sigma\cdot \br * V +T\cdot \sigma V=0,
%\label{kernel}
%\end{eqnarray} 
%where the dimensionless variable $s$ has its range $[-1/2,1/2]$
%and $T_i= - s\theta U \delta_{i3}$ for a monopole. 
%The Higgs field can be obtained in terms of $V$ by
%\begin{eqnarray}
%\phi= U\int^{1/2}_{-1/2} ds\, s\, V^\dagger * V\,.
%\label{higgs}
%\end{eqnarray}
%Rewritten
%in terms of dimensionless variables, the kernel equation becomes
%\begin{eqnarray}
%\left[-{d\over d(s U\sqrt\theta )} + 
%\sigma_1 \left(\tilde{x}-{i\over 2}{\partial \over
%\partial \tilde{y}}\right)
%+\sigma_2 \left(\tilde{y}+{i\over 2}{\partial \over
%\partial \tilde{x}}\right)
%-U\sqrt{\theta} \left(s-{z\over \theta U}\right)
% \right] V=0,
%\label{kernel2}
%\end{eqnarray} 
%where $\tilde{x}=x/\sqrt{\theta}$ and  
%$\tilde{y}=y/\sqrt{\theta}$.
%It is now  clear
%that the scale governing %x and y directional size is 
%the $z$ directional size is $\theta U$.

%
\begin{figure}[tb]
\epsfxsize=3.0in
%\centerline{
\hspace{.8in}\epsffile{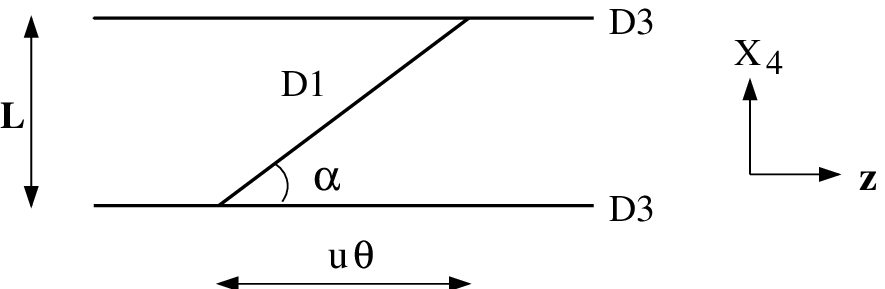}
%}
%\vspace{.1in}
\\
{\small Figure~2:~The brane configuration of a noncommutative 
monopole where $L=u\, l_s^2$ and $\cot\alpha= u\theta/L$.}
\end{figure}

This size $L_z$ measures the extension of this Dirac string along
z-direction.  The distance between two D3-branes is given by $L= u\,
l_s^2$ with the string length scale
$l_s=\sqrt{2\pi\alpha'}$. Consequently the tilting angle $\alpha$ is
\begin{eqnarray}
 \alpha = \tan^{-1} l_s^2/\theta\; ,
\label{slope}
\end{eqnarray} 
which does to zero in the zero slope limit. 
% Fig. 2 illustrates the
%brane configuration corresponding to the noncommutative BPS monopole.
Since the length of monopole or D-string is then $\sqrt{L^2+L_z^2}$,
the tension of the monopole is 
\begin{eqnarray}
T_0= {M\over \sqrt{L^2+(u\theta)^2}}= {2\pi \over 
g^2_{\rm YM}}
{q_m\over 
\sqrt{l_s^4 +\theta^2}} \, .
\label{slope2}
\end{eqnarray} 
In particular, in the zero slope limit of $2\pi \alpha'\rightarrow 0$,
the tension becomes identical to the tension of the Dirac string
$M/(u\theta)= {2\pi q_m \over g^2_{\rm YM}\theta}$, which agrees with
the value in \cite{gross}. (In the full field theory, this should be
true when $u$ is very large and so the Dirac string is very
long. Then the field energy of two $U(1)$ monopoles can be negligible
in comparison.)

The solution of a moving monopole can be generated from the static
solution. Let $\bar{A}_\mu(x,y,z;\theta)$ and
$\bar\phi(x,y,z;\theta)$ be the solution describing a static monopole
for arbitrary $\theta$.  For monopole moving on the $x-y$ plane, we
restrict our discussion to ${\bf v}=v \hat{x}$ due to the rotation
symmetry
around the z axis. The corresponding  Lorentz transformation is given by
\begin{eqnarray}
t'=\gamma(t-vx)\,, \ \ x'=\gamma (x-vt),\  y'=y,\  z'=z\,.
\label{lorentz}
\end{eqnarray} 
It is now straightforward to verify that 
the the solution of a moving monopole   is 
\begin{eqnarray}
A'_\mu (t,x,y,z;\theta)&=&{\partial {x'}^\nu\over \partial {x}^\mu} 
\,\,\bar{A}_\nu \Bigl(\gamma(x-vt),y,z;\gamma\theta\Bigr)\nonumber\\
\phi' (t,x,y,z;\theta)&=& 
\bar{\phi} \Bigl(\gamma(x-vt),y,z;\gamma\theta\Bigr)\,.
\label{movemono}
\end{eqnarray}
The effective size in the $z$ direction is given by $L_z(v)=u\theta
\gamma = \gamma L_z$. Since the effective noncommutative scale
$\theta_{\rm eff}$ is given by $\gamma \theta$, the monopole looks
more tilted when it is moving on $x-y$ plane. Namely, the distance
$L=u\, l_s^2$ between D3-branes unchanged but the Dirac string
connecting $U(1)$ monopole and $U(1)'$ anti-monopole gets `elongated'
by the factor $\gamma$, and so the size becomes $L_z(v)=\gamma
u\theta$.  The tilted slope angle is now given by $\alpha(v) =
\tan^{-1} ( l_s^2/\gamma\theta)$. Thus, in the relativistic speed, the
$U(2)$ monopole would look like a very long stick.
A few comments are in order. First, it should be noted that
the full $U(2)$ solution is not known thus far for finite $\theta$.
Hence we do not have the explicit solution for the 
finite velocity either.
Secondly, the shape of monopole is governed by the noncommutativity scale 
$\theta$ together 
with the scale $L_z$ that controls
the dipole structure. The whole deformation of the monopole shape is 
partly from
the change of the dipole structure as the scale $L_z$ changes
effectively.
There is also deformation from the Lorentz contraction
and the change in the effective noncommutativity scale. 
The latter part of deformation is universal to
all moving noncommutative solitons including the case of 
scalar noncommutative solitons.

The moving solution to the $z$ direction is obtained similarly.
The Lorentz boost transformation in the z direction reads 
\begin{eqnarray}
t'=\gamma(t-vz)\,, \ \ x'=x,\  y'=y,\  z'=\gamma (z-vt)\,.
\label{lorentz2}
\end{eqnarray} 
The the corresponding moving solution is given by
\begin{eqnarray}
A'_\mu (t,x,y,z;\theta)&=&{\partial {x'}^\nu\over \partial {x}^\mu} 
\,\,\bar{A}_\nu \Bigl(x,y,\gamma(z-vt);\theta\Bigr)\nonumber\\
\phi' (t,x,y,z;\theta)&=& 
\bar{\phi} \Bigl(x,y,\gamma(z-vt);\theta\Bigr)\,,
\label{movemono2}
\end{eqnarray}
and the effective noncommutative scale remains unchanged and only
ordinary Lorentz contraction in the $z$ direction by the factor
$1/\gamma$ has occurred.  The image of the $U(2)$ monopole in the three
space has the size $L_z(v)=u\theta/\gamma$ and titling angle becomes
$\alpha(v)=\tan^{-1}(\gamma l_s^2/\theta)$.

Contrary to the case of the scalar noncommutative soliton,
the energy and momentum of the moving monopole behaves
like a massive particle. Namely, they are respectively
given by $E=\gamma M$ and ${\bf P}=\gamma M {\bf v} $, which 
may be checked directly  by inserting the above solutions to
 definitions of 
the energy and the momentum.

Finally let us consider the case of dyons or 
$(p_e,q_m)$-strings. The dyons satisfy BPS equations
\begin{eqnarray}
B&=& \cos\xi \, D\phi\nonumber\\
E&=& \sin\xi \, D\phi\,.
\label{dyonbps}
\end{eqnarray}
We define electric charge by
\begin{eqnarray}
p_e ={1\over  g^2_{\rm YM} \, u} \int_{r=\infty} dS_k \tr E_k \phi,
\label{echarge}
\end{eqnarray} 
and, for the elementary excitations of W-bosons,
it takes integer values as expected. Since $E$ and $B$ are related through
the angle $\xi$
in the above BPS equations, the ratio of 
the electric charge $p_e$ and the magnetic charge $q_m$ is then found to be
\begin{eqnarray}
{p_e\over q_m}= {2\pi\over g^2_{\rm YM}} \tan\xi \,  .
\label{pqrelation}
\end{eqnarray} 
%
%The effective noncommutativity scale can be understood
%from
%\begin{eqnarray}
%\theta\cos^2\xi {\partial \over \partial (x\cos\xi)} a({\bf r})
% {\partial \over \partial (y\cos\xi)} b ({\bf r})=
%\theta {\partial \over \partial x} a({\bf r})
% {\partial \over \partial  y} b ({\bf r})\,.
%\label{dyonsolution2}
%\end{eqnarray}
For a given magnetic solution, the corresponding dyon solution can be
found by a scale transformation and
a Lorentz boost in the extra dimension if we view as
$\phi=A_4$. In the noncommutative case, one should take into account
the change of the effective noncommutative scale. The corresponding
dyon solution satisfying the above BPS equations are
\begin{eqnarray}
&& A_i = \bar{A}_i({\bf r}\cos\xi;\theta\cos^2\xi) \nonumber \\
&& \phi = \bar{\phi}({\bf r}\cos\xi;\theta\cos^2\xi) \nonumber \\
&& A_0 = \sin\xi \; \bar{\phi}({\bf r}\cos\xi; \theta\cos^2\xi) 
\end{eqnarray}
The energy is  determined again by surface integral and 
can be given in terms of charges by
\begin{equation}
M_{p_e,q_m} = {2\pi\over g^2_{\rm YM}} {q_m u\over \cos\xi}=
u \sqrt{ \left( \frac{2\pi q_m}{g^2_{YM}}\right)^2 +p_e^2}
\, .
\label{dyonmass}
\end{equation}
%If $L$ were the length of (p,q) dyon, the the tension would be
%${1\over \l_s^2} \sqrt{p^2+q^2/\hat{G}^2_s}$ but this is not the case 
%because the $(p,q)$ string is tilted. 
The length scale of the image in $z$ direction is shrunken to $L^D_z =
u\theta /\sqrt{1+ (g^2_{YM} p_e/2\pi q_m)^2}$ and the tilting angle
changes to  $\alpha_D=\tan^{-1}\Bigl( (l_s^2/\theta)\sqrt{1+ (g^2_{YM}
p_e/2\pi q_m)^2}\Bigr)$.    In the zero slope limit, the
string tension for the Dirac string becomes
\begin{eqnarray}
T_{(p_e,q_m)}=  \frac{2\pi q_m}{g^2_{YM}\theta} \left( 
1+ (g^2_{YM} p_e/2\pi q_m)^2 \right)\, , 
\label{dyontension2}
\end{eqnarray}
for $q_m\neq 0$.  When $p_e=0$, the tension becomes $T
%={q\over
%\hat{G}_s\theta}
= {2\pi q\over g^2_{\rm YM}\theta}$, which agrees with
the value given previously.  When $q_m=0$, Eq.~(\ref{dyontension2}) is
not valid.  Indeed the fundamental string tension goes to infinite in
the zero slop limit. When moving, 
these $(p_e,q_m)$ dyonic configurations would
also go through the same elongation or contraction as the pure
monopole configuration.

In this note, we have observed the shape of moving noncommutative
solitons is elongated.  In short, it accentuates the UV/IR mixing. As
the velocity approaches the light velocity, the transverse size grows
indefinitely, which is a phenomena residing in the IR regime of the
theory.  Moving monopoles in the noncommutative SYM theories can have
a similar elongation, following  the change of tilting of D-strings
connecting D3 branes.  We have obtained the tension and tilting of
static $(p_e,q_m)$-dyons.

Investigation of the free motion is the first step toward
understanding of dynamical characteristics of noncommutative solitons.
For the more detailed dynamics, further studies are required on the
moduli dynamics of noncommutative solitons.  Especially, the quantum
moduli dynamics of the false vacuum bubble in the noncommutative
scalar field theory will be of interest. Also our observation on
monopoles would also apply to the noncommutative open string theories
studied recently~\cite{ncos}.

\noindent{\large\bf Acknowledgment} We would like to 
thank 
P. Yi and C. Zachos  for
enlightening discussions and R. Jackiw for pointing out 
 Ref.~\cite{hobart}.
  D.B. would like to thank H. S.  Yang for
valuable comments. This work is supported in part by KOSEF 1998
Interdisciplinary Research Grant 98-07-02-07-01-5 (DB and KL) and by
UOS Academic Research Grant (DB). K.L. thanks the hospitality in Niels
Bohr Institute where the part of work is done.

\end{document}